\documentclass[aps,prd,nofootinbib,showpacs,superscriptaddress,preprint]{revtex4}
\usepackage{graphicx}
\usepackage{epsfig}
\usepackage{amsmath}
\usepackage{amsfonts}
\usepackage{amssymb}

\def\spose#1{\hbox to 0pt{#1\hss}}
\def\simlt{\mathrel{\spose{\lower 3pt\hbox{$\mathchar"218$}}
     \raise 2.0pt\hbox{$\mathchar"13C$}}}
\def\simgt{\mathrel{\spose{\lower 3pt\hbox{$\mathchar"218$}}
     \raise 2.0pt\hbox{$\mathchar"13E$}}}
\begin{document}

\title{\large \sc Quantum Structure of Geometry: Loopy and
fuzzy?}

\author{\sc Alejandro Corichi}\email{corichi@matmor.unam.mx} 
\affiliation{\it 
Instituto de Matem\'aticas - Morelia,
Universidad Nacional Aut\'onoma de M\'exico, 
UNAM-Campus Morelia, A. Postal 61-3, Morelia, Michoac\'an 58090,
Mexico.
}
\affiliation{Institute for Gravitational Physics and Geometry,
Physics Department, Pennsylvania State University, University Park
PA 16802, USA}

\author{\sc Jos\'e A. Zapata}\email{zapata@matmor.unam.mx}
\affiliation{\it 
Instituto de Matem\'aticas - Morelia,
Universidad Nacional Aut\'onoma de M\'exico, 
UNAM-Campus Morelia, A. Postal 61-3, Morelia, Michoac\'an 58090,
Mexico.
}

\begin{abstract}
In any attempt to build a quantum theory of gravity, a central issue is to
unravel the structure of space-time at the smallest scale. 
Of particular relevance is the possible definition of coordinate functions within the theory and the study of their algebraic properties, such as non-commutativity. 
Here we approach this issue from the perspective of loop quantum
gravity and the picture of quantum geometry that the formalism offers.
In particular, as we argue here, this emerging picture has two main elements: i) The nature of the quantum geometry at Planck scale is one-dimensional, polymeric with
quantized geometrical quantities and; ii) Appropriately defined operators corresponding to coordinates by means of intrinsic, relational, constructions  become non-commuting. This particular feature of the operators, that operationally localize points on space, gives rise to an emerging geometry that is also, in a precise sense, fuzzy.
\end{abstract} 

\maketitle

The question of what the microstructure of space-time is has been
in the minds of many theoretical physicists for the past 80 years
(if not more), ever since it was clear to Einstein in 1916 that we
need a grander theory that unites gravity and the quantum: a
quantum theory of gravity. We do not yet have such a satisfactory
framework that brings together the quantum and gravitational
realms. However, there are questions that can be posed even in the
absence of a final theory. For instance, one can ask whether there
is a minimal length scale in the theory (See \cite{Garay} for a
review).   Another question that comes about is the issue of the
structure of space-time at this scale. The nature of this question
is however, a bit less precise since the question itself depends
heavily on the details of the approach one is considering. For
instance, if one assumes the existence of space-time (as a
$4$-dimensional background manifold) as a-priory given, one will
pose different questions as one would in purely canonical
approaches where the theory is defined on a 3-manifold $\Sigma$
and where a $4$-dimensional space-time emerges only in some
semi-classical regime.

One particular feature of the underlying structure of quantum
space-time is a possible non-commutativity at the fundamental
scale. Somewhat surprisingly, even in this case, there seems to be
no consensus regarding what this means exactly. Sometimes it is
assumed that the {\em coordinates} will be non-commutative while,
alternatively, non-commutativity could be understood in the sense
of Connes' non-commutative geometry \cite{Connes}, where not even
the notion of a manifold survives. More recently, a pile of
results coming from different approaches hint to a possible
non-commutative structure at the fundamental scale of the theory.
This has been extensively explored in the literature on string
theory, gauge fields and in other frameworks, so we will not
review it here. Some of these approaches suffer, we believe, from
a serious drawback: 
While their objective is to understand the micro structure of 
space-time in the attempt to go down to the most basic object on 
which physics is constructed, their starting point is not a 
fundamental object. 
General relativity taught us more
than 90 years ago that coordinates are not fundamental objects for
the description of the physics, and any approach that uses them as
a starting point would seem to contradict the lessons of GR. Once
the dynamical objects, described in terms of fields (gravitational
included) have been prescribed, then one could construct a notion
of coordinates defined dynamically from the theory itself (such as
the so-called GPS coordinates \cite{gps}). Taking this lesson
seriously implies that any notion of non-commutative coordinates
as a starting point faces the burden of proving itself  consistent
with background independence.

In this note we shall advocate a different perspective on the
structure of spacetime; a picture that has arisen from the
framework known as {\em loop quantum gravity} (LQG) \cite{lqg}. To
begin with, as presently understood, loop quantum gravity is a
theory defined on a canonical setting, which means that there is
no 4-dimensional object  we can call space-time, but only a
background 3-manifold $\Sigma$ where the theory is defined.
Space-time becomes a derived object. There is no background metric
$q^{\rm o}_{ab}$ defined on $\Sigma$, so the quantum theory has to
be defined, in this sense, metric background-independent. What
makes LQG radically different from other canonical approaches to
quantum gravity (such as its predecessor, Quantum Geometrodynamics
in terms of the Wheeler-DeWitt equation), is that it can actually
be well defined. More precisely, we have for the first time a
mathematically consistent framework where some of the basic
objects, such as a Hilbert space and finite operators, exist (For
details see \cite{lqg} and for pedagogical introductions see
\cite{AA:2004} and \cite{playa}). A crucial input in the
construction is the choice of objects that are rendered as basic
for the quantization of the theory, namely connections and loops.
Connections are present because Einstein's theory was recast as a
gauge theory with a Yang-Mills structure; loops are important
because they are the natural objects to use when defining simple
functions of the connection, namely, parallel transports and
holonomies. The metric information on $\Sigma$ is contained in the
triads $e^a_i$ that are the other objects containing the full
information of the phase space. Any 3-geometric object is
constructed out of this canonical variable.

An important aspect of the formalism is that, apart from the
choice of variables, one does {\it not} presuppose the nature of
the microstructure of space nor its properties. There is no input
about discreteness at any scale. One literally lets the theory
lead.

{}From these basic objects, one can construct quantum operators that
represent geometrical quantities such as areas  of surfaces
$\hat{A}[S]$ and volumes of regions $\hat{V}[R]$. The operators
are finite and suffer from no regularization infinities. They can
be rigourously regulated without the need of any renormalization.
With them one can explore the nature of the quantum geometry that
Loop Quantum Gravity offers for us. The first thing that the
formalism tells us is that the quantum (metric) geometry is
concentrated on lower dimensional objects. Loops (or more
generally, edges of graphs) carry ``fluxes of area", very much in
the style of Faraday, while vertices of the graph are responsible
for the volume of regions. The second important feature is that
the spectra of the operators are {\em discrete}, providing in a
precise sense a {\em quantization} of the geometry. Thus, if
$|\Psi_{\Upsilon,\vec{j}}\rangle$ is a quantum state labelled by a
graph $\Upsilon \in \Sigma$ and a set of labels $\vec{j}$, one for
each edge, then the action of the operator $\hat{A}[S]$ on
$|\Psi_{(\Upsilon,\vec{j})}\rangle$ is given by,
\[
\hat{A}[S]\cdot|\Psi_{(\Upsilon,\vec{j})}\rangle=\left(
8\pi\,\gamma\,\ell^2_{\rm pl}\;\sum_{v_i} \sqrt{j_i(j_i+1)}
\right)
|\Psi_{(\Upsilon,\vec{j})}\rangle\, ,
\]
where the sum is over all the intersection points $v_i$ between
$S$ and $\Upsilon$, and the (half integer) $j_i$ corresponds to
the label of the corresponding edge. In particular, note that there
is a minimum area gap in the theory corresponding to a single
intersection and spin $j=1/2$: $A_{\rm min}=4\pi\gamma\ell^2_{\rm
pl}\,\sqrt{3}$. The real number $\gamma >0$ is a free parameter of
the theory that needs to be fixed by some `experiment' (it is the
analogue of the $\theta$ parameter in QCD). The best (at present)
proposal for fixing the value of the parameter comes from Black
Hole physics and it is of the order of $\gamma=0.274\ldots$. With
this choice the minimum area gap $A_0\approx 6\,\ell^2_{\rm pl}$
is of the order of the Planck area. This is the smallest area
possible and as we have already emphasized, it {\em is} a
prediction of the theory.

The third, and somewhat unexpected feature, is that the loop
quantum geometry is inherently non-commutative: operators
associated to intersecting surfaces $S_1$ and $S_2$ do {\em not}
commute:
\[
[\hat{A}[S_1], \hat{A}[S_2]]\neq 0\, .
\]

For more details see \cite{non-comm,q-arena}. This, of course
implies that there is an intrinsic uncertainty in the geometry,
and therefore an induced {\em fuzziness} in the spatial quantum
geometry. Any attempt of localization by means of operators
associated to surfaces will be subject to a basic Heisenberg
uncertainty $(\Delta A[S_1])(\Delta A[S_2]))\neq 0$. Many
questions come to mind regarding the non-commutativity of these
fundamental geometric operators. In particular, one may wonder how
we can recover an apparently smooth, commutative geometry on
large scales? How does this affect the algebraic nature of
symmetry groups, such as the Lorentz group, that are so useful in
describing physics at low energies? Does this non-commutativity
survive in the full dynamical setting?, etc.

So far, we only have partial answers to some of these questions,
but the picture of the geometry of space that LQG presents for us
is quite novel and intriguing. Let us now explore some of these
new aspects of this quantum description of geometry at the
fundamental scale. For instance one question that is basic to any
further consideration is the following: In what sense can we
regard individual points $p_j$ of the manifold $\Sigma$ as endowed
with a physical meaning? In other words, does the concept of a
point on space even make sense? In the context of classical
general relativity we know the answer: given that the theory is
invariant under diffeomorphisms, the concept of a point as an 
abstract entity dissolves. Instead what makes sense is the point
not as an abstract mathematical object but as the location where
matter fields and gravity have some particular property (for
instance the point where two word-lines intersect, or the point
where light is emitted by a source, etc.). Nevertheless, any point
on the manifold is as good as any other point given that all
fields are smooth objects and are thus `well defined' on any point
of space. What is then the situation in the quantum realm? After
all, quantum fields are a very different class of objects than its
classical counterparts.

Is loop quantum geometry point-less? Let us start to answer that
question by considering states $|\Psi\rangle_{\rm diff}$ that are
solutions to the (gauge) and spatial-diffeomorphism constraint of
the theory. In terms of the objects
$|\Psi_{\Upsilon,\vec{j}}\rangle$, they can be obtained by means
of an {\em average} over the orbits of the diffeomorphisms acting
on the state. In a sense the original graph $\Upsilon$ is no
longer localized on the manifold, but is `smeared out by means of
the average' procedure. What matters of the state
$|\Psi\rangle_{\rm diff}$ are the self relational properties of the
original state $|\Psi_{\Upsilon,\vec{j}}\rangle$. That is, the
incidence relations of the graph $\Upsilon$, the labels  $j_i$ on
the edges, etc. It should be clear then that it is only the graph
and points on it that can be given any physical meaning. There are
simply no points `outside the graph and on the manifold $\Sigma$'.
The manifold $\Sigma$ itself has dissolved. The next observation
is that within a graph (that can be seen as abstract) there 
is a clear distinction between points on it. Thus, points that
lie on an edge are qualitatively different from those that are
vertices (with more than two incident edges). In terms of the
geometric operators with which we can probe the state (and thus
determine what is the quantum geometry defined by the state), a
point $p_e$ on an edge can carry a flux of area (detected by
acting on the state with a surface intersecting the graph
$\Upsilon$ at the point $p_e$), but it corresponds to vertices
$p_v$ on $\Upsilon$ the privilege of endowing 3-dimensional
regions $R$ with a volume; the corresponding volume operator only
`excites' vertices with valence four and higher. Otherwise, the
contribution to the volume is zero. Furthermore, when coupling any
kind of matter, it is again the vertices with volume the ones that
contribute by providing the volume
for the matter Hamiltonian. Thus, if one is to define the meaning
of `points of space' in a relational manner, by means of dynamical
matter fields, it will only be vertices of the graph that will be
elevated to the level of {\em physical points}. 

Having established this fact, one could try to define the notion
of {\em coordinates of physical points} and study their properties. For
instance, 
if we define in a {\em relational} manner, using only
objects defined on the graph and geometrical quantities, operators
$\hat{Q_i}$ assigning  coordinates to physical points, 
it turns out that the resulting operators are {\em non-commuting}.
%
Let us now explain how one can define this type of operators 
(the choice is by no means unique). The underlying principle behind these constructions is that the coordinates have to be defined intrinsically from the dynamical objects of the theory. In this sense they become relational.

Let us first consider the situation for  the definition of these
coordinate functions in the classical scenario. 
Our construction will use a fiducial metric $q_o$ on $\Sigma$, 
and a set of reference points $(p_0,p_1,p_2,p_3)$ to construct the 
coordinate functions (functions on phase space). 
We will 
use the fiducial metric and the 
reference points to construct, given any point $p$
`near $p_i$', three 2-spheres $S_i$ as follows: The sphere $S_1$
for example will be the $q_0$-sphere that has $(p_0,p_2,p_3,p)$ on it.
That is, it is the (unique, provided some tame conditions are
satisfied) sphere defined by these four points. In the same
manner, we can define $S_2$ by the four points $(p_0,p_1,p_3,p)$,
and $S_3$ by $(p_0,p_1,p_2,p)$. Now we define the coordinates,
\[
Q_i:=A[S_i]
\]
That is,  the coordinates 
are given by the value of the area
corresponding to the $S_i$ sphere, as given by the metric on 
$\Sigma$ defined by the phase space point (not the fiducial metric). 
In a neighborhood of the reference points, we are able to
define the spheres and thus the coordinates $Q_i$. As in the
definition of any coordinate chart, the choices of the reference
points and the fiducial metric are totally arbitrary. Any other
choice will, of course give different coordinates for the same
point. We again remark that these coordinates are 
`$q$-coordinates' in the
sense they depend on the phase space point (in this case, the
spatial metric).

Let us now try to export the construction to the quantum realm.
First we have to specify the reference points $(p_0,p_i)$. From
the discussion before we can only choose vertices on the graph
$\Upsilon$ (physical points) 
as the reference points and furthermore, the only
points $p$ for which the coordinates will be defined are also physical 
points. 
In the construction we will make use of the
fiducial metric 
to select the surfaces $S_i$; as we have mentioned for the classical 
coordinates, this is
part of the arbitrary specification of any coordinate system. 
The second observation is that we
shall define the corresponding operators in the kinematical
setting, where the graphs are embedded into $\Sigma$. 
%
The objects to be `quantized' are the area functions that now will be
replaced by area operators as,
\[
\hat{Q}_i:=\hat{A}[S_i]
\]
The important thing to note is 
that the point $p$ corresponds to the point where all
the spheres $S_i$ intersect, and given that $p$ is a physical
point, it is a node with al least valence four in the
graph. This is precisely the condition for the area operators not
to commute. Thus, in general, we have that,
\[
[\hat{Q}_i,\hat{Q}_j ]\neq 0
\]
The coordinate operators associated to physical 
points are non-commuting. Needless to say, we
have employed a rather simple prescription,  so our results are
not exhaustive. In the small sample of `slightly more elaborate
constructions' that we have explored, we have also found that the
corresponding quantum coordinates do not commute \cite{q-arena}.

In a situation where matter fields have been coupled to
loop quantum gravity, we could define similar coordinate operators whose
basic building blocks are the area operators $\hat{A}[S_i]$, but this 
time the surfaces $S_i$ would be specified using
dynamical fields. It is exactly this situation what we had in mind when we
demanded that the point to be assigned coordinates
and the reference points had to be {\em physical points}
(nodes of the graph that specifies the state).
These coordinate operators would be background
independent and they would still not commute. 

The construction so far was somewhat kinematical since we employed a 
particular fiducial metric and an embedding of the graph into the
manifold $\Sigma$. Can we elevate our construction to the diffeo 
invariant realm? That is, without the need for a particular embedding? 
The answer is in the affirmative but the construction is somewhat subtle.
In our construction of coordinate operators we used both the fiducial metric 
and the embedded graph that defines the {\it q-arena} we are considering:
the quantum geometry defined by the graph $\Upsilon$ and the state 
$\Psi_\Upsilon$ defined on it. The fiducial metric allowed us to construct the auxiliary quantities such as the spheres defined by the reference points and the
point to be measured. But once this objects are specified and the operators 
$\hat{Q}_i$ are defined, they only know about `combinatorial' aspects of the state:
the vertex on which they act and the particular action on the incoming edges to that vertex. We can extend the action of these operators to states obtained by the action of a diffeomorphism $\phi$: simply maintain the same action on the `internal' part of the state, {\it for all states on the orbit of the diffeomorphism group}.
Defined in such a way, the operators become well defined on the space of
diffeo-invariant states. This procedure can be seen as the equivalent, on the quantum realm, of the ``evolving constants of the motion" prescription on the classical 
side where a gauge invariant function is defined as constant along the orbits of the
constraint functions.   

The scenario here presented is by no means conclusive, but certainly these
results hint to a geometry at the Planck scale that is not only
polymeric, but also intrinsically non-commutative. This represents
the first example of concrete non-commutativity arising as a {\em
consequence} of a well defined mathematical framework that unites
the gravitational and quantum realms.\footnote{Perhaps the most notable exception
are the intriguing results by Freidel and collaborators \cite{freidel}} 
Needless to say, the story
is far from over and the subject deserves further attention. We hope to report
the results of such investigations in a near future \cite{q-arena}.

Apart from the possible non-commutativity of space-time, and just
as general relativity changed so drastically our notions of space
and time, a final theory of quantum geometry will for sure bring a
new and deeper perspective to these notions. What we can all
probably agree upon is that is are still far from a definite
answer.

We thank J.M. Reyes for discussions. This work was partially supported by  CONACyT U47857-F and NSF PHY04-56913 grants, by the Eberly
Research Funds of Penn State.

\end{document}